# Photospheric Abundances: Problems, Updates, Implications

H. Holweger

*Institut für Theoretische Physik und Astrophysik, Universität Kiel, 24098 Kiel, Germany*

**Abstract.**
Current problems encountered in the spectroscopic determination of photospheric abundances are outlined and exemplified in a reevaluation of C, N, O, Ne, Mg, Si, and Fe, taking effects of NLTE and granulation into account. Updated abundances of these elements are given in Table 2.

Specific topics addressed are (1) the correlation between photospheric matter and C I chondrites, and the condensation temperature below which it breaks down (Figure 1), (2) the question whether the metallicity of the Sun is typical for its age and position in the Galaxy.

## INTRODUCTION

The Anders-Grevesse (1989) abundance table [1] has become a widely used reference for solar-system abundances. Follow-up compilations have appeared, the most recent one we are aware of being that by Grevesse and Sauval (1998) [2].

Most, if not all of the photospheric abundance determinations compiled rely on two basic approximations commonly made in abundance work on solar-type stars: local thermodynamic equilibrium (LTE) and onedimensional photospheric models.

Meanwhile deviations from LTE can be taken into account through detailed modeling of radiative and collisional processes, although important atomic data are still lacking (see below). Apart from NLTE abundance corrections, elaborate 2D and 3D hydrodynamical models of stellar convection coupled with radiative transfer are becoming available. It is the aim of the present contribution to provide a small data base for a few key elements that permits to assess the magnitude of the effects of NLTE and solar granulation in photospheric abundance determinations.

## THE COMPLEXITY OF PHOTOSPHERIC LINE FORMATION

Judged from the simple Gaussian shape of most photospheric absorption lines, the solar photosphere might be regarded as an easy-to-model absorption tube. In reality it is a highly inhomogeneous and dynamic plasma permeated by an intense, anisotropic radiation field whose spectral energy distribution differs from that of a black body. It is not at all evident, although commonly assumed, that local thermodynamic equilibrium (LTE) should hold, i.e. that quantum-mechanical states of atoms, ions, and molecules are populated according to the convenient relations of Boltzmann and Saha, valid strictly only in thermodynamic equilibrium. Fortunately one can model this plasma with NLTE calculations, taking the essential microscopic processes into account. Such simulations allow to identify the processes that drive the plasma away from LTE, as well as those that act to thermalize it. It has become clear that departures from LTE arise from the above mentioned deviations from black-body radiation via bound-bound and bound-free radiative transitions, while the main thermalizing processes in solar-type stars are inelastic collisions with electrons and hydrogen atoms. Unfortunately accurate cross-sections are scarce. For the potentially dominating hydrogen collisions only a coarse estimate is available (see, e.g. [3]).

Absorption-line spectroscopy requires counting photons that were removed from the continuous spectrum by the species of interest. However, the picture of a 'reversing layer' that absorbs an underlying continuum is much too simple. Re-emission of both continuum and line photons occurs at all heights, although absorption generally dominates if temperature decreases outward. In order to determine elemental abundances one needs to know these processes in all atmospheric layers that contribute to the formation of the continuous and line spectrum. This implies knowledge of temperature and other variables as a function of height. In addition, horizontal and vertical fluctuations associated with convection and waves have, in principle, to be taken into account.

Another complication arises if the line considered is strong enough to show effects of saturation, i.e. the linear relation between equivalent width and the product of $f$-value and abundance breaks down. The deviation from linearity depends on line-broadening mechanisms that have traditionally introduced substantial uncertainties in stellar spectroscopy: Doppler broadening by nonthermal velocity fields ('microturbulence'), mostly of convective origin, and broadening by collisions with ambient particles, primarily hydrogen atoms and electrons. A more detailed discussion of these problems may be found e.g. in [3].

# UPDATED PHOTOSPHERIC ABUNDANCES OF SELECTED ELEMENTS

In this section we discuss and reevaluate recent data available for a few important elements including iron and oxygen.

Departures from LTE in the solar photosphere will be taken into account as outlined in [4], employing model atoms described in the respective sections. The HM photospheric model [5] is adopted as a default. This model has been used in many of the abundance determinations on which the above mentioned compilations are based. To illustrate the model-dependence of the results, calculations for iron and oxygen have also been carried out with the VAL model [6]. The latter is widely used as a reference model for chromospheric studies. However, its photospheric structure fails to reproduce the excitation equilibrium of temperature sensitive molecules like OH [7] [8] [9] and CO [10]. The NLTE abundance corrections are summarized in Table 2.

As a novel feature this analysis also considers the effects of photospheric temperature inhomogeneities associated with convection. Representing the mean vertical temperature structure by the empirical HM model, abundance corrections are applied to account for horizontal temperature inhomogeneities associated with granulation, following the approach described by Steffen [11], based on 2D numerical models of solar granulation and LTE line formation. For the purpose of this study abundance corrections were kindly made available by M. Steffen (private communication). They are summarized in Table 2.

We note that effects of NLTE and granulation are treated here as independent, second-order effects rather than attempting an ab-initio approach that includes everything. The smallness of the effects (Table 2) is consistent with this differential approach, which has the advantage of being computationally affordable and allowing to assess the importance of either effect in the Sun and in other solar-type stars.

In the following sections individual elements are discussed in a somewhat unconventional order, starting with iron and oxygen, two important cases where a more detailed report will be given.

## Iron

In the Anders-Grevesse abundance table [1] the recommended photospheric iron abundance is $\log N_{\mathrm{Fe}} = 7.67 \pm 0.03$. In the recent compilation by Grevesse and Sauval [2] a value of $\log N_{\mathrm{Fe}} = 7.50 \pm 0.05$ is quoted. Note that the abundance has changed significantly, and the assessment of error limits is now more conservative.

Intriguingly, error limits quoted in the literature are defined in two different ways. The more optimistic version is the error of the mean abundance (derived from individual lines), while the standard deviation of individual lines is more conservative. Both differ by a factor $\sqrt{n}$, where n is the number of lines involved. The entry for iron (and for many other elements) in the Anders-Grevesse abundance table [1] conform to the former prescription, while the more conservative version is used throughout the present paper. In this version errors of individual lines are not assumed to be independent, with the reasoning that errors depending on wavelength, equivalent width, or excitation cannot be excluded.

The recommended iron abundance is based exclusively on lines of ionized iron. In the solar photosphere iron occurs mainly as Fe II, hence these lines are much less sensitive to the thermal structure of the photospheric model than Fe I lines are. In addition, a number of independent sources of $f$-values are available, both experimental and theoretical. Previous Fe II analyses [12] [13] have consistently inferred a lower iron abundance than that adopted in the Anders-Grevesse compilation, $\log N_{\mathrm{Fe}} = 7.48$. Recently, Schnabel et al. [14] (SKH99) have used improved lifetime measurements for Fe II levels in a new abundance analysis based on the line list of [12], inferring an even lower value, $7.42 \pm 0.09$. The quoted error limits conform to the more conservative prescription mentioned above. The mean abundance following from all 13 Fe II lines in Table 2 of SKH99 (entering the three infrared lines with half weight) is 7.419 with a standard deviation of 0.082.

For each of the 13 Fe II lines NLTE abundance corrections $\Delta \log N = \log N_{\mathrm{NLTE}} - \log N_{\mathrm{LTE}}$ have been calculated using a model atom described in [15]. NLTE effects in Fe II lines are extremely small, $|\Delta \log N_{\mathrm{Fe}}| \leq 0.001$, and mean LTE and NLTE abundances agree to within 0.001 dex. The photospheric iron abundance, including NLTE effects, but still without granulation corrections, becomes $\log N_{\mathrm{Fe}} = 7.419 \pm 0.082$.

If the HM model is replaced by the VAL model, the mean abundance increases by 0.049 dex both in LTE and NLTE.

The effect of granulation on the iron abundance has been determined for a small number of Fe II lines representative of the SKH99 sample, leading to a mean correction of +0.029 dex. The recommended photospheric iron abundance is then obtained by combining the (1D) LTE result with corrections for effects of NLTE (1D) and granulation (2D): $\log N_{\mathrm{Fe}} = 7.448 \pm 0.082$.

## Oxygen

The photospheric abundance listed in the Anders-Grevesse table [1] is $\log N_{\mathrm{O}} = 8.93 \pm 0.035$, while Grevesse and Sauval [2] recently have recommended $\log N_{\mathrm{O}} = 8.83 \pm 0.06$. Again, a significant change in the preferred abundance and a trend towards more conservative error limits is to be noted.

The oxygen abundance inferred here is derived from atomic lines, while the Anders-Grevesse value is based on an unpublished analysis of the CNO group using molecular lines. Like Fe II, O I is the dominant species, while molecules such as OH are trace constituents whose formation is quite temperature sensitive. Grevesse & Sauval [2] report CNO abundances which they have derived from molecular as well as atomic lines. Unfortunately no details are given and the authors regard the result (quoted above) as preliminary. Previous analyses based exclusively on atomic lines [16] [17] have led to closely coincident abundances of 8.86 and 8.87, respectively. However, this agreement is somewhat surprising in view of the different sets of lines and $f$-values that have been used. Moreover in [16] LTE is assumed and the HM model is used, while a detailed NLTE analysis is carried out in [17] in combination with a flux-constant ATLAS6 model. We feel that this situation needs clarification and have reevaluated the O I data as follows.

The oxygen lines analyzed are listed in Table 1. Our sample is based on the line list given in [18], supplemented by four additional infrared lines from [16]. In order to minimize blend problems, lines with uncertain profiles were omitted because they are likely to be contaminated by other atomic or molecular species whose contribution, if not properly accounted for, will inevitably lead to fictitiously high oxygen abundances. Specifically, all three lines marked ':' in Table 5 of [18] have been rejected ($\lambda\lambda 5577.3$, 6156.8, and 6363.8 Å) together with the multiplet at 8446 Å which is strongly perturbed by Fe I. Our selection is in accordance with [19]. Of the infrared lines listed in [16] $\lambda 9262.8$ Å is not accepted because its profile is badly blended. In our line formation calculations of multiplets 67 and 64 fine structure splitting is taken into account. In both cases the combined equivalent width of the multiplet is quoted in Table 1, and only one abundance value is assigned to each triplet. Equivalent widths refer to the center of the solar disk. All $\log gf$ values were taken from the NIST data base [20]; most of them are based on quantum-mechanical calculations [21]. LTE abundances derived from individual lines are presented in Table 1. Assigning equal weight to all lines, a mean LTE abundance of 8.780 results, with a standard deviation of 0.071.

Our LTE abundance is 0.08 dex lower than that inferred by Biémont et al. [16]. The line abundances quoted in Table 3 of [16] permit to trace the reason for this difference. Most of it is due to lines we have not used because of less reliable profiles; if the four lines in question ($\lambda\lambda 6156.8$, 8446.3/8446.8 and 9262.8 Å) are excluded from the Biémont et al. sample and the remaining eight lines in their Table 3 given equal weight, the mean abundance decreases to 8.803. The residual 0.023 dex difference can be attributed to slightly different $f$ values and equivalent widths. The four perturbed lines, if averaged separately, lead to a 0.120 dex higher abundance, illustrating the bias towards higher abundances that may arise if blends are included in abundance determinations.

The blend problem has also been addressed by Reetz [22] in the context of the two well-known [O I] lines ($\lambda\lambda$ 6300.3/6363.8 Å). These lines have been entered with high weight in many solar and stellar abundance determinations since they were considered as safe with respect to NLTE effects. While this is certainly true (see Table 1), their extreme faintness makes them very sensitive to blends. Indeed, Reetz argues that $\lambda 6363.8$ (which we have not used) is more strongly perturbed by a CN line than realized previously, and that $\lambda 6300.3$ may contain contributions by a Ni I line. This could explain why $\lambda 6300.3$ yields the highest abundance of our sample. Detailed spectrum synthesis of the [O I] line regions using all available data bases may help to solve this problem.

NLTE calculations for oxygen have been carried out employing a model atom described in [23]. Deviations from LTE in O I are most important for the strongest lines, amounting to $\Delta \log N_{\mathrm{O}} = -0.065$ for $\lambda 7771.9$ Å (Table 1). The mean NLTE correction is $\Delta \log N_{\mathrm{O}} = -0.028$. In all NLTE calculations inelastic collisions with H atoms have been taken into account, assuming a scaling factor $S_{\mathrm{H}} = 1.0$ in accordance with Takeda [17]. The photospheric oxygen abundance, including NLTE effects but without granulation corrections, becomes $\log N_{\mathrm{O}} = 8.752 \pm 0.078$.

The LTE and NLTE results quoted refer to the HM photospheric model. If the VAL model is adopted instead, the LTE abundance increases by 0.072 dex. NLTE effects are hardly different; the average NLTE abundance correction is now $\Delta \log N_{\mathrm{O}} = -0.027$ (VAL) instead of $-0.028$ (HM). Thus the NLTE abundance derived from the VAL model is 0.073 dex higher than the corresponding HM result.

**TABLE 1.** Photospheric oxygen lines used for abundance analysis

| Wavelength (Å) | Mult. | E.P.(eV) | log gf | W (mÅ) | log $N_{O,LTE}$ | log $N_{O,NLTE}$ | $\Delta_{gran}$ |
|---:|---:|---:|---:|---:|---:|---:|---:|
| 6158.15 | 67 | 10.741 | -1.841 | ... | ... | ... | ... |
| 6158.17 | 67 | 10.741 | -0.996 | ... | ... | ... | ... |
| 6158.19 | 67 | 10.741 | -0.409 | 5.0 | 8.780 | 8.775 | $-0.085$ |
| 6300.30 | 92 | 0.000 | -9.774 | 4.3 | 8.921 | 8.921 | $+0.015$ |
| 7771.94 | 56 | 9.146 | 0.396 | 88.0 | 8.834 | 8.769 | $+0.015$ |
| 7774.17 | 56 | 9.146 | 0.223 | 71.0 | 8.776 | 8.714 | $+0.007$ |
| 7775.39 | 56 | 9.146 | 0.001 | 54.0 | 8.741 | 8.694 | $+0.000$ |
| 9265.83 | 64 | 10.741 | -0.719 | ... | ... | ... | ... |
| 9265.93 | 64 | 10.741 | 0.126 | ... | ... | ... | ... |
| 9266.01 | 64 | 10.741 | 0.712 | 35.6 | 8.718 | 8.695 | $-0.028$ |
| 11302.4 | 63 | 10.741 | 0.076 | 14.0 | 8.697 | 8.684 | $-0.033$ |
| 13164.9 | 76 | 10.989 | -0.033 | 16.3 | 8.772 | 8.766 | $-0.017$ |

Granulation corrections for the individual O I lines are given in the last column of Table 1. Note that corrections may have either sign, depending on excitation and line strength. On average, abundances derived from conventional 1D models have to be adjusted by $-0.016$ dex if the sample of Table 1 is adopted for analysis. The recommended photospheric oxygen abundance is the combination of the LTE result with corrections for effects of NLTE and granulation: $\log N_O = 8.736 \pm 0.078$.

## Carbon

The photospheric abundance listed in the Anders-Grevesse table [1], $\log N_C = 8.56 \pm 0.04$, is based on the same analysis of CNO-group molecular lines used also for oxygen. Grevesse and Sauval [2] suggest $\log N_C = 8.52 \pm 0.06$ as a preliminary result of an unpublished revised evaluation of both molecular and atomic species.

As mentioned above, molecular lines are quite temperature sensitive. Our recommended value is based on published analyses of C I lines [24] [25]. The LTE carbon abundances derived from Table 2 of [24] and Table 2 of [25] are $8.583 \pm 0.133$ and $8.600 \pm 0.098$, respectively. The detailed NLTE calculations of the former paper [24] have shown that in C I NLTE effects depend strongly on equivalent width, and a mean NLTE correction of $-0.05$ dex was derived from this sample. For the sample of [25] we find a mean NLTE correction of $-0.04$ dex. Combining the respective LTE abundances and NLTE corrections and taking the average of both determinations leads to a NLTE abundance of $\log N_C = 8.571 \pm 0.108$.

Granulation corrections were determined for one representative line of either sample rather than for the dozens of lines. Its wavelength, excitation potential, and equivalent width were chosen as the mean value of each of the two samples. This resulted in two sets of line parameters, (9800 Å, 7.8 eV, 90 mÅ) for the sample of [24], and (12700 Å, 8.7 eV, 83 mÅ) for the [25] sample. The corresponding granulation corrections are $+0.023$ and $+0.019$ dex, respectively (M. Steffen, private communication).

The final photospheric carbon abundance is derived by combining the respective LTE abundances and corrections for NLTE and granulation, and taking the average of both samples. The recommended value is $\log N_C = 8.592 \pm 0.108$.

## Nitrogen

The photospheric abundance listed in the Anders-Grevesse table [1] is $\log N_N = 8.05 \pm 0.04$ has resulted from the same (unpublished) analysis of CNO-group molecular used also for C and O. Grevesse and Sauval [2] suggest a significantly lower value, $\log N_N = 7.92 \pm 0.06$, derived from a preliminary analysis of atomic and molecular lines.

In analogy to carbon and oxygen our recommended value is based on a NLTE analysis of photospheric N I lines [15]. Table 2 of [15] leads to a mean NLTE correction of $-0.032$ dex and a NLTE abundance of $\log N_N = 8.049 \pm 0.097$. The $f$-values employed were taken from the Opacity Project data base. In the meantime a NIST compilation has appeared [20], which is used here. On average the new NIST $\log gf$ values are 0.048 dex larger than the OP values (for the line sample of [15]), implying a downward readjustment of the nitrogen abundance by the same amount for a given set of equivalent widths. Omitting the highly discrepant line at 10575.9 Å whose $f$-value is obviously in error, the new NLTE abundance becomes $\log N_N = 8.001 \pm 0.111$.

As in the case of carbon, granulation effects are derived for a 'mean' N I line with parameters representative of the sample used. The mean line is characterized by (9600 Å, 11.1 eV, 3.8 mÅ), implying that the lines are all weak, and of high excitation. Line of this type are quite sensitive to temperature inhomogeneities [11]. In the case of N I the granulation correction amounts to $-0.070$ dex. Taking this into account leads to the recommended photospheric abundance of $\log N_\mathrm{N} = 7.931 \pm 0.111$.

## Magnesium

Identical entries for magnesium are found in the abundance tables of Anders-Grevesse [1] and Grevesse-Sauval [2], $\log N_\mathrm{Mg} = 7.58 \pm 0.05$. This value dates back to the 1984 review of Grevesse [26].

Our recommended value is based on an analysis of Mg II lines [27] because this one-electron system permits accurate quantum-mechanical calculations of $f$-values, while this is notoriously difficult for Mg I. The LTE abundance [27] is $\log N_\mathrm{Mg} = 7.54 \pm 0.06$. NLTE calculations for magnesium in the Sun [28] have focused on Mg I. The resulting departures from LTE are very small throughout the photosphere. Since magnesium is mostly present as Mg II, we expect even smaller departures for this dominant species. Hence zero NLTE corrections are assumed for the solar magnesium abundance, which is based exclusively on Mg II lines. This is supported by existing NLTE calculations for Mg II in A-type stars. Even though the radiation field of an A star with $T_\mathrm{eff}$ = 9500 K deviates from blackbody radiation much more than that of the Sun, NLTE corrections are typically only $-0.04$ dex [29].

The granulation correction for a representative Mg II line with mean parameters (8800 Å, 9.4 eV, 45 mÅ) was determined to be $-0.002$ dex. We note that this is not typical for high-excitation Mg II lines in general, but is due to cancellation of the dependence of granulation effects on excitation and on line strength. The recommended photospheric abundance is $\log N_\mathrm{Mg} = 7.538 \pm 0.060$.

## Silicon

Like Mg, the Si abundance $\log N_\mathrm{Si} = 7.55 \pm 0.05$ listed in the Anders-Grevesse [1] and Grevesse-Sauval [2] tables was adopted from the 1984 review by Grevesse [26]. It is based on the LTE analysis of Si I and Si II by Becker et al. [30].

Recently Asplund [31] has carried out a re-analysis using a 3D hydrodynamical model. The result, $\log N_\mathrm{Si} = 7.51 \pm 0.04$, includes effects of granulation but rests on the assumption of LTE.

In an independent parallel study of essentially the same line sample Wedemeyer [32] has derived an LTE abundance of $7.560 \pm 0.066$ in close agreement with the abundance tables, using a conventional 1D model. His study also presents detailed NLTE calculations. For the combined Si I/II line sample an average NLTE correction of $-0.010$ dex is derived from standard 1D modeling, thus the NLTE result (without granulation effects) becomes $7.550 \pm 0.056$. The smaller standard deviation in NLTE as compared to LTE is mainly due to the good fit of Si I and Si II achieved by the NLTE calculations.

Wedemeyer [32] quotes a granulation correction of +0.021 dex, derived by Steffen [11] in the same way as the corrections for the other elements. The corresponding LTE abundance including granulation correction, is $7.581 \pm 0.066$. which may be compared with Asplund's [31] value of $7.51 \pm 0.04$. Although both values agree within their mutual error limits, the 0.07 dex difference is somewhat disturbing and deserves further investigation.

We adopt the mean of both determinations and assume the same NLTE correction of $-0.010$ dex for both analyses. The recommended photospheric abundance thus becomes $\log N_\mathrm{Si} = 7.536 \pm 0.049$.

## Neon

The abundance listed in the Anders-Grevesse table [1] is $\log N_\mathrm{Ne} = [8.09 \pm 0.10]$; Grevesse and Sauval [2] recommend $\log N_\mathrm{Ne} = [8.08 \pm 0.06]$. Square brackets in the tables indicate that these values were not derived from photospheric absorption lines. Although Ne I does have lines in the visible part of the spectrum, these lines are of very high excitation and hence too weak to be observable in the solar photosphere.

The Anders-Grevesse value is based on solar particle and local galactic data, while Grevesse and Sauval have adopted the more recent result obtained by Widing (1997) [33] from EUV spectroscopy of emerging active regions.

We prefer the latter source over solar particle data because, as Widing [33] argues, emerging flux events most likely permit direct observation of unfractionated photospheric material. The abundance value of 8.08 was derived in [33] from the Ne/Mg ratio following from EUV data by combining it with the Anders-Grevesse photospheric Mg abundance of 7.58. We make two modifications: (1) we use also oxygen, in addition to magnesium, as the photospheric reference element, and (2) adopt the updated photospheric values derived above.

The abundances following from emerging active regions are quoted in Table 4 of [33] in terms of abundance ratios, Ne/Mg $= 3.16 \pm 0.07$ (based on five observations) and O/Ne $= 6.75 \pm 0.65$ (two observations). The Ne/Mg ratio then translates into a photospheric neon abundance of 8.038 while the O/Ne ratio yields 7.907. We take the mean, assigning weight 5 to the former and weight 2 to the latter. The recommended photospheric abundance becomes $\log N_{Ne} = 8.001 \pm 0.069$.

**TABLE 2.** Recommended photospheric abundances ($\log N_H = 12$)

| Element | log N | $\Delta_{NLTE}$ | $\Delta_{gran}$ | $\delta_{AG89}$ | $\delta_{GS98}$ |
|---|---|---|---|---|---|
| 6 C | 8.592 ± 0.108 | −0.045 | +0.021 | +0.03 | +0.07 |
| 7 N | 7.931 ± 0.111 | −0.032 | −0.070 | −0.12 | +0.01 |
| 8 O | 8.736 ± 0.078 | −0.028 | −0.016 | −0.19 | −0.09 |
| 10 Ne | 8.001 ± 0.069 | ... | ... | −0.09 | −0.08 |
| 12 Mg | 7.538 ± 0.060 | ∼ 0 | −0.002 | −0.04 | −0.04 |
| 14 Si | 7.536 ± 0.049 | −0.010 | +0.021 | −0.01 | −0.01 |
| 26 Fe | 7.448 ± 0.082 | +0.000 | +0.029 | −0.22 | −0.05 |

## Update Summary

The updated abundances are summarized in Table 2. Effects of NLTE and photospheric granulation are included in the abundances, and specified separately in the next two columns. Also given are the differences between the new abundances and those compiled by Anders and Grevesse [1] ($\delta_{AG89}$) and Grevesse and Sauval [2] ($\delta_{GS98}$) in the sense 'new minus old'.

## SOLAR AND METEORITIC ABUNDANCES

The close correlation between solar and meteoritic abundances of non-volatile elements is well known. The unique mineralogy of CI chondrites qualifies their matter as an essentially unaltered condensate of the solar nebula. One may ask if there is a well-defined condensation temperature below which the compositional correlation breaks down.

This question is addressed in Figure 1. The quantity plotted is $\log(X/Si)_{CI} - \log(X/Si)_\odot$, the depletion of various elements in CI chondrites with respect to the solar photosphere. Included are the elements up to the iron group and two volatile heavier elements, Cd and Pb. For meteorites we have adopted data from [1], for the Sun those listed in Table 2 supplemented by data from [2]. Not included are B, F, and Cl because their abundance must be regarded as highly uncertain (F and Cl are based on sunspot spectra). For the strongly depleted elements (hydrogen and noble gases) upper limits are given. Not shown is lithium, which is depleted in the Sun by 2.2 dex due to thermonuclear destruction.

Approximate condensation temperatures have been collected from the literature; actual values will depend on the conditions under which condensation occurs. Nevertheless, Figure 1 shows a strikingly close correlation down to the condensation temperature of Pb and Cd, the most volatile condensed elements with reliable photospheric data. For Cd, $T_C \approx 420$ K [34]. Even oxygen ($T_C \approx 180$ K [35]) obviously has condensed to a remarkable degree. It seems that the solar nebula was well-mixed at the time when bulk of the CI matter condensed from the gas phase, which is believed to have occurred at a distance of several AU from the central star.

## THE SUN AND THE METALLICITY OF NEARBY STARS

Is the metallicity of the Sun typical for stars of solar mass and age? This question is addressed in a recent discussion by Gustafsson [36] of the comprehensive set of data on bright F and G main sequence stars presented by Edvardsson et al. [37]. Gustafsson emphasizes that their sample was chosen to be biased towards low metallicities: the stars were selected to be evenly distributed over the metal abundance range $-1.0 > [Me/H] > +0.3$. In other words, compared to a volume limited sample the number of metal deficient stars was enhanced, hence the more metal-rich stars like the Sun appear less numerous. Furthermore, we note that the selection cutoff near $[Me/H] = +0.3$ of the Edvardsson et al. sample will cause an additional thinning out of the sample towards the metal-rich side. All this obviously has led some authors to believe that the age-metallicity relation depicted in Figure 14a of [37] implies that the Sun is metal-rich by about 0.2 dex. However, if the sample is corrected for the low-metallicity bias (which is not trivial) and restricted to nearby stars with ages between 4 and 5 Gyr, the solar 'anomaly' reduces to $[Fe/H] = -0.09 \pm 0.22$ [36].

The Edvardsson et al. [37] analysis is strictly differential in the sense that the same instrument was used for recording stellar and solar (flux) spectra, and the same suite of model atmospheres was employed for analysis. Another strictly differential analysis was carried out by Fuhrmann [38]. Again no attempt was made to establish a volume limited sample, but unlike [37] no metallicity cutoff just above the solar level was imposed. Figure 11 of [38] suggests that [Fe/H] as well as [Mg/H] of the Sun is quite typical for nearby disk dwarfs.

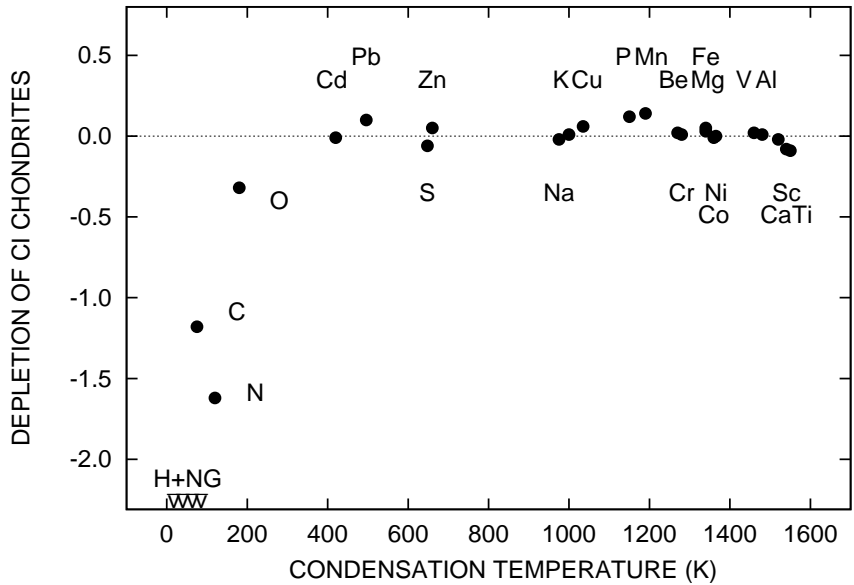

**FIGURE 1.** Comparison of abundances in CI chondrites with those in the solar photosphere

By contrast, another recent study of the age-metallicity relation [39] is not strictly differential but makes use of a metallicity calibration of *uvby* photometry [40] whose main source of abundance data is a compilation [41] which dates back to 1983 and includes a variety of earlier stellar and solar abundance determinations. While the clear trend of metallicity with age shown in Figure 13 of [39] is not affected by this uncertainty, the zero point of the [Fe/H] scale - which apparently indicates that the Sun is anomalously metal-rich - must be regarded as preliminary.

Probably the most accurate data set currently available is a by-product of the search for extrasolar planets. A sample of 77 single G dwarfs was used by Butler et al. [42] for comparison with planet-bearing stars. It represents a volume-limited set of field stars whose photometric metallicities were calibrated with strictly differential spectroscopic analyses. Figure 7 of [42] clearly shows that the Sun is about 0.1 dex more iron rich than nearby field stars (and that stars with detected planets are even more metal-rich). Since the age of the solar neighborhood is more than twice the age of the Sun [43], an enhanced solar metallicity fits qualitatively into the picture of galactic chemical evolution, although the metal enrichment during $\sim 5$ Gyr predicted by most models is larger than 0.1 dex. Thus the Sun would appear to be metal-poor for its age, rather than metal-rich! Possibly the concept of a strict age-metallicity relation has to be revisited. A similar conclusion has been reached very recently [44] in view of the detection of super-metal-rich, $\sim 10$ Gyr old stars.

## THE SUN AND GALACTIC ABUNDANCE GRADIENTS

Does the Sun fit into the galactic abundance gradients determined from various nonstellar and stellar sources? While a general review of this topic is beyond the scope of this paper, we briefly mention some consequences of the updated solar abundances listed above (Table 2).

We may take advantage of the large body of observational data on galactic H II regions, planetary nebulae, and B stars recently compiled by Hou et al. [45]. In their Figure 6 - which may serve as a basis for our discussion - abundances of various elements are plotted versus galactocentric distance $R_G$, showing a rather large scatter and, in most cases, a more or less pronounced abundance gradient. The scatter possibly reflects real variations from object to object as well as uncertainties in the determinations. In any case it permits to assess whether the Sun ($R_G = 8.5$ kpc) fits into the overall picture emerging from recent data.

Close inspection of Figure 6 of [45], with the updated abundances of C, N, O, Ne, Mg, and Si taken into account, shows that the Sun is quite typical, even in the case of oxygen whose previous solar value seemed higher than typical. However, the solar carbon abundance appears to be $\sim 0.4$ dex higher compared to that derived from galactic B stars. It is highly improbable that the solar value is in error by this large amount. This important element deserves further study.

## SUMMARY AND OUTLOOK

The elemental composition of the solar photosphere closely resembles that of type CI meteorites down to a condensation temperature of 400 K and agrees within 0.1 dex with that of stars in our galactic neighborhood. This qualifies solar-system abundances as an excellent reference for galactic and extragalactic studies.

Spectroscopic observations of photospheric matter can be carried out with high precision, yet the interpretation in terms of abundances is inevitably much more indirect than counting particles. While photon spectroscopy will remain indispensable for stars other than the Sun, the measurement of solar energetic particles and the solar wind - in combination with the determination of hydrogen and with a better understanding of fractionation processes - will hopefully lead to an improved knowledge of the isotopic and elemental composition of the solar photosphere, comparable in accuracy to that already achieved for meteorites.

## ACKNOWLEDGMENTS

The author is grateful to Matthias Steffen for providing granulation abundance corrections in advance to publication. Support by ESA and by the Physikalisches Institut, University of Bern, is gratefully acknowledged.